\documentclass[12pt]{iopart}
\usepackage{iopams}

\usepackage{dsfont}
\usepackage{graphicx}
\usepackage[colorlinks=true, pdfborder={0 0 0}, citecolor=blue,linkcolor=black]{hyperref}
\usepackage{cite}
\usepackage{doi}
\usepackage{cleveref}

\usepackage{etoolbox}
\patchcmd{\numparts}{\addtocounter{equation}{1}}{\refstepcounter{equation}}{}{}

\usepackage{braket}
\newcommand{\ac}[1]{#1_{\mathrm{ac}}}
\newcommand{\del}{\partial}
\newcommand{\bn}{\bi{n}}

\begin{document}

\setcounter{page}{1} 

\title{Thermodynamically consistent model of an active {O}rnstein-{U}hlenbeck particle}

\date{\today}%
\author{Jonas H. Fritz$^1$ and Udo Seifert$^1$}
\address{$^1$ II. Institut für Theoretische Physik, Universität Stuttgart, 70550 Stuttgart, Germany}
\ead{useifert@theo2.physik.uni-stuttgart.de}
\begin{abstract}
{Identifying the full entropy production of active particles is a challenging task. We introduce a microscopic, thermodynamically consistent model, which leads to active Ornstein-Uhlenbeck statistics in the continuum limit. Our minimal model consists of a particle with a fluctuating number of active reaction sites which contribute to its active self-propulsion on a lattice. In addition, the model also takes ordinary thermal noise into account. This approach allows us to identify the full entropy production stemming from both thermal diffusion and active driving. Extant methods based on the comparison of forward and time-reversed trajectory underestimate the physical entropy production when applied to the Langevin equations obtained from our model. Constructing microscopic Markovian models can thus provide a benchmark for determining the entropy production in non-Markovian active systems.}
{}{}
\end{abstract}

\maketitle 

\section{Introduction} \label{sec:introduction}
The research of active matter \cite{marchetti_hydrodynamics_2013,bechinger_active_2016,ramaswamy_active_2017} has explored many facets for which thermodynamic properties play an important role. Examples include heat engines driven by active matter and their design principles \cite{krishnamurthy_micrometre-sized_2016,pietzonka_autonomous_2019,holubec_active_2020,fodor_active_2021,gronchi_optimization_2021,datta_second_2022,majumdar_exactly_2022} and the onset of motility induced phase separation \cite{buttinoni_dynamical_2013,cates_motility-induced_2015,ro_model-free_2022}. In most of this theoretical work, the underlying active matter models for particles can be grouped into three distinct categories with different characteristic motion. Run and tumble particles (RTP)\cite{schnitzer_theory_1993,tailleur_statistical_2008} perform straight motion at a constant speed in conjunction with Poisson-distributed tumble events in which the particles change their orientation.
Active Brownian particles (ABP)\cite{romanczuk_active_2012,chaudhuri_active_2014} move at a constant velocity in the direction of a director, which changes its orientation continuously. Finally, active Ornstein-Uhlenbeck particles (AOUPs)\cite{berthier_nonequilibrium_2014,szamel_self-propelled_2014,martin_statistical_2021} additionally feature a stochastically changing velocity, described by an Ornstein-Uhlenbeck process. 

Fundamentally, all of these active matter models are driven out of equilibrium by an active reservoir such that they break time-reversal symmetry. In addition, the state of the particle is not only given by its position, but also by its orientation, which makes the dynamics of the former non-Markovian if the orientation is not accessible. 

This property has made it difficult to straightforwardly apply the methods of stochastic thermodynamics \cite{sekimoto_stochastic_2010,jarzynski_equalities_2011,seifert_stochastic_2012,van_den_broeck_ensemble_2015} to investigate thermodynamic quantities such as the physical entropy production of these systems \cite{speck_stochastic_2016,gnesotto_broken_2018,szamel_stochastic_2019,obyrne_time_2022}. In turn, attempts have been made to find new measures for the departure from equilibrium on the level of Langevin trajectories. These, however, do not have a straight-forward interpretation as physical entropy production, but rather measure the irreversibility by comparing the probability of forward and backward trajectory \cite{ganguly_stochastic_2013,fodor_how_2016,mandal_entropy_2017,speck_active_2018,dadhichi_origins_2018,shankar_hidden_2018} or by only identifying dissipated heat \cite{marconi_heat_2017}. Constructing the backward trajectory is ambiguous however, since the time-reversal for the active noise history can be chosen to be even or odd. For an in-depth discussion about the role and intricacies of time-reversal, and different interpretations of the resulting pseudo-entropy productions, we point to the reviews in \cite{caprini_entropy_2019,fodor_irreversibility_2022}. Another way to approach this issue is given in \cite{dabelow_irreversibility_2019,dabelow_irreversibility_2021}, where the colored noise history is integrated out of the path weight and encoded in a memory kernel.

In addition, exact expressions for the entropy production of active particles have been derived on the level of microscopic Markovian models, to which the principles from stochastic thermodynamics are easily applied. So far, this has been done for ABPs \cite{pietzonka_entropy_2017} and for RTPs \cite{padmanabha_fluctuations_2023}. For active Ornstein-Uhlenbeck particles, however, such a model still does not exist to the best of our knowledge. The aim of this paper is to fill this gap and to provide a minimal, thermodynamically consistent model which leads to the equations of motion of an AOUP assumed \textit{a priori} in other work. The paper is structured as follows: In \cref{sec:model} we introduce the model in one dimension and calculate the exact expression for the mean entropy production rate. In \cref{sec:comparison}, we then compare this expression to results found previously. Finally, in \cref{sec:ddim}, we show that our model also exhibits AOUP statistics in higher dimensions and conclude in \cref{sec:conclusion}.

\section{One-dimensional model} \label{sec:model}
\subsection{Definition}
\begin{figure}
  \centering
  \includegraphics[width=\textwidth]{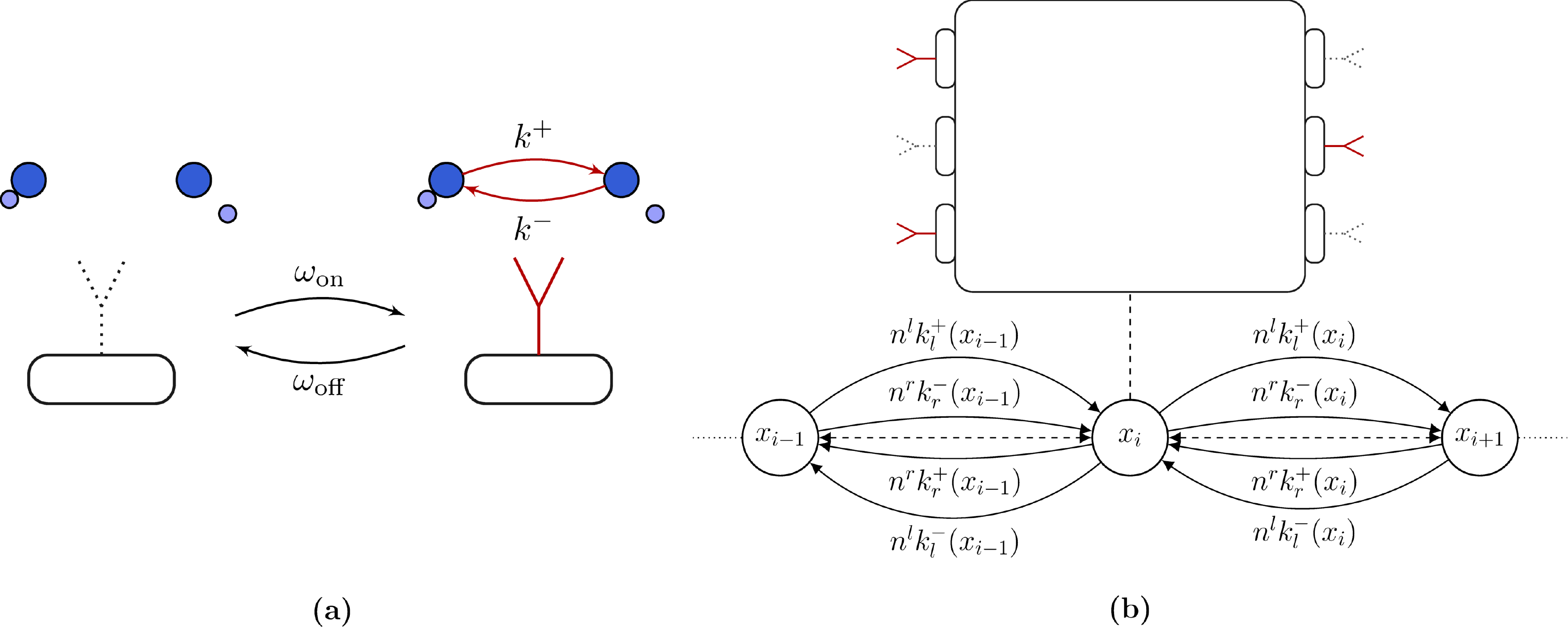}
  \caption{Illustration of the active driving mechanism. (a) A single reaction site is either inactive (dotted and gray) or active (red). In the active state, the reaction site catalyzes some fuel-consuming chemical reaction with rate $k^+$, which releases chemical free energy $\Delta\mu$. Thermodynamic consistency requires that the reverse reaction with rate $k^-$ is also possible. The blue circles symbolize the reactants. Switching between the active and inactive state occurs with rates $\omega_{\mathrm{on, off}}$. (b) The particle with internal degree of freedom $(n^l,n^r)$ at position $x_i$ on the one dimensional lattice. On the left side of the particle, $n^l=2$ sites are active, and on the right one $n^r=1$ site is active. Each active site on the left pushes the particle to the right if the fuel-consuming reaction with rate $k^+_l(x_i)$ takes place and pulls it to the left if the fuel-generating reaction with rate $k_l^-(x_{i-1})$ occurs. Reactions on the right side of the particle have the opposite effect, pushing it to the left with rate $k_r^+(x_{i-1})$ and pulling it to the right with rate $k_r^-(x_i)$. In addition, the particle can diffuse thermally, indicated by the dashed arrows between the states.}
  \label{fig:sketch}
\end{figure}
We start by considering the model in one dimension, where the particle is on a discrete lattice with lattice constant $a$ and sites $x_i$. The motion of the particle can be driven in either direction by some process, like a chemical reaction, which draws its energy from the environment, e.g., from a chemical reservoir. This process has a forward rate $k^+$, which extracts free energy from the reservoir, and a backward rate $k^-$, which puts it back into the reservoir. 
These rates are parametrized by their asymmetric part 
  \begin{equation}
    k^+/k^-=\mathrm{e}^{\beta\Delta \mu}\, ,
  \end{equation}
where $\Delta \mu$ is the free energy drawn from the reservoir with each forward reaction. We set the inverse temperature $\beta=1$ from here on and measure entropy in units of $k_{\mathrm{B}}$.

As shown in figure$\,\,$\ref{fig:sketch}, the reaction takes place at phosphorylation sites at the surface of the particle, either on the right side of the particle, where the forward reaction drives it to the left and the reverse reaction to the right, or, \textit{vice versa}, on the left side of the particle. Whatever the exact underlying mechanism is, thermodynamic consistency requires that the reverse process is also possible.

On each side of the particle, there are $N$ reaction sites in total, each of which is in either of two states. They can be active with probability $p_A$, or inactive with a probability $p_I=1-p_A$. Only active sites can catalyze reactions, inactive sites do not contribute to the driving. The number of active reaction sites on the left and the right are $n^l$ and $n^r$ respectively. The activation of an inactive site occurs with rate $\omega_{\mathrm{on}}$ and the deactivation of an active one with rate $\omega_{\mathrm{off}}$, which are given by
\begin{equation}
  \omega_{\mathrm{on}} = \kappa \sqrt{p_A/p_I} \quad \mathrm{and}\quad \omega_{\mathrm{off}} = \kappa \sqrt{p_I/p_A}\,,
  \label{eq:def:omega}
\end{equation}
where $\kappa$ sets the timescale at which these transitions take place. Thus, the dynamics of the internal state of the particle, characterized by $(n^l,n^r)$, follows the master equation 
\begin{eqnarray}
  \partial_t p(n^l,n^r,t) &=& \omega_{\mathrm{on}}\left[(N-n^l+1) p(n^l-1,n^r,t)\right.\cr
  &\phantom{}&\left. + (N-n^r+1) p(n^l,n^r-1,t)\right] \cr
  &\phantom{}& + \omega_{\mathrm{off}}\left[(n^l+1) p(n^l+1,n^r,t) + (n^r+1) p(n^l,n^r+1,t)\right] \cr
  &\phantom{}& - \left[(n^l+n^r)(\omega_{\mathrm{on}}+\omega_{\mathrm{off}})\right] p(n^l,n^r,t)\, .
\end{eqnarray}
Each active site contributes to the mean velocity and mean diffusion of the particle. We call this characteristic mean velocity per reaction site $u_0$ and the characteristic mean diffusion coefficient per site $D_0$, which are defined as
\begin{equation}
	u_{0} \equiv  (k^+ - k^-)a\quad \mathrm{and}\quad  D_{0} \equiv  (k^+ + k^-)a^2/2 \,.
\end{equation}
For fixed $(n^l,n^r)$, the particle will then move with the average velocity $\langle\dot{x}|n^l,n^r\rangle=\left(n^l-n^r\right)u_0$ and diffuse as $\langle (x(t)-x(0))^2|n^l,n^r  \rangle=2\left(n^l+n^r\right)D_0 t$. With these definitions we can write 
  \begin{equation}
    k^\pm = D_{0}/a^2 \pm u_{0}/2a\, .
  \end{equation}
If there is an additional external spatial potential $V(x)$, thermodynamic consistency requires
  \begin{equation}
    k^\pm_l (x_i) =  k^\pm \exp\left\{-\left[V(x_i)-V(x_{i+1})\right]/2\right\}
  \end{equation}
  for the rates of reaction sites on the left side and 
  \begin{equation}
    k^\pm_r (x_i) =  k^\pm \exp\left\{-\left[V(x_{i+1})-V(x_{i})\right]/2\right\}\,,
  \end{equation}
  for the reaction sites on the right side. This distinction takes into account that the forward and backward direction drive the particle in different directions of the potential depending on the side of the particle at which they take place.

\subsection{Dynamics}
Turning to the dynamics of the particle, we introduce the active current for a fixed number of active reaction sites $n^l$ and $n^r$ as
  \begin{eqnarray}
      \ac{j}(x_i,n^l,n^r) &\equiv& j_{x_i,x_{i+1}}(n^l,n^r)\cr
       &=&  n^l \left[p(x_i,n^l,n^r) k^+_l(x_i)-p(x_{i+1},n^l,n^r)k^-_l(x_{i})\right]\cr
        &\phantom{}&+n^r \left[p(x_i,n^l,n^r)k^-_r(x_{i})-p(x_{i+1},n^l,n^r)k^+_r(x_{i})\right]\cr
        &=&\ac{j}^l(x_i,n^l,n^r) + \ac{j}^r(x_i,n^l,n^r)\, .
        \label{eq:jac}
  \end{eqnarray} 
The first term, $\ac{j}^l(x_i,n^l,n^r)$, accounts for the current generated by the $n^l$ active reaction sites on the left side of the particle and has two contributions: The particle can be pushed from $x_i$ to $x_{i+1}$ by a forward reaction with rate $k^+_l(x_i)$ and can be pulled from $x_{i+1}$ to $x_i$ by the fuel synthesizing backward reaction with rate $k^-_l(x_i)$. The second term, $\ac{j}^r(x_i,n^l,n^r)$, accounts for the $n^r$ active reaction sites on the right side of the particle, where the reaction with rate $k^+_r(x_{i})$ displaces it to the left, and the reverse reaction with rate $k^-_r(x_i)$ displaces it to the right, as is illustrated in figure$\,\,$\ref{fig:sketch}. Preparing for the continuum limit, where we treat $x$ as a continuous variable and let $a\rightarrow 0$, we expand the rates in $a$ as 
  \begin{equation}
    k^\pm_l(x_i) \approx (D_0/a^2 \pm u_0/2a) \left[1\mp  a\del_{x} V(x_i)/2\right]\,,
    \label{eq:k_l}
  \end{equation}
  and 
  \begin{equation}
    k^\pm_r(x_i) \approx (D_0/a^2 \pm u_0/2a) \left[1\pm  a\del_{x} V(x_i)/2\right]\,,
    \label{eq:k_r}
  \end{equation}
  as well as the stationary distribution
  \begin{equation}
    p(x_i+a) \approx p(x_i)\left[1+ a\del_{x} \mathrm{ln}\,p(x_i)\right]\,,
  \end{equation}
  and obtain in leading order $1/a$
 \begin{numparts}\label{eq:1d_curr}
  \begin{eqnarray}
  \ac{j}^l(x_i,n^l,n^r) &=& a^{-1}p(x_i,n^l,n^r) \left\{ n^l u_0\right.\cr
  &\phantom{}&\left.-D_0n^l\left[\del_x V(x_i)+\del_x\mathrm{ln}p(x_i,n^l,n^r)\right]\right\}\,,\\
  \cr
  \ac{j}^r(x_i,n^l,n^r) &=& a^{-1}p(x_i,n^l,n^r) \left\{ -n^r u_0\right.\cr
  &\phantom{}&\left.-D_0n^r\left[\del_x V(x_i)+\del_x\mathrm{ln}p(x_i,n^l,n^r)\right]\right\}\,,
    \end{eqnarray}
\end{numparts}
  such that the total current becomes 
  \begin{eqnarray}
    \ac{j}(x_i,n^l,n^r)&=&a^{-1}p(x_i,n^l,n^r) \left\{ n^- u_0\right.\cr
    &\phantom{}&\left. -D_0n^+\left[\del_x V(x_i)+\del_x\mathrm{ln}p(x_i,n^l,n^r)\right]\right\}\,.
    \label{eq:1d_curr_tot}
  \end{eqnarray}
  We have also introduced 
  \begin{equation}
    n^\pm \equiv n^l\pm n^r\,,
    \label{eq:defnpm}
  \end{equation}
  as a shorthand. Describing the internal state through $n^-$ and $n^+$ as independent degrees of freedom is equivalent to describing it via $n^l$ and $n^r$. Since the physical expressions below will depend on the former description, these variables will prove more convenient.
  \subsection{Entropy production}
  The entropy production rate arising from the active motion is given as usual in stochastic thermodynamics as \cite{seifert_stochastic_2012,van_den_broeck_ensemble_2015}
  \begin{equation}
    \ac{\sigma} = \sum_{i,n^l,n^r}\left\{ \ac{j}^l(x_i,n^l,n^r)\mathrm{ln}\left[\frac{k_l^+(x_i) }{ k_l^-(x_i) }\right] +\ac{j}^r(x_i,n^l,n^r)\mathrm{ln}\left[\frac{k_r^-(x_i) }{ k_r^+(x_i) }\right]\right\} \, ,
  \end{equation}
with the currents given in leading order $1/a$ by (\ref{eq:1d_curr}). It is crucial to distinguish whether a jump occurred due to a reaction on the left or right side of the particle, as each has different contributions to the entropy production. By expanding the rates as in (\ref{eq:k_l}) and (\ref{eq:k_r}), we obtain 

\begin{numparts}
\begin{eqnarray}
    \mathrm{ln}\left[\frac{k_l^{+}(x_i)}{k^{-}_l(x_i)}\right]&\approx& \frac{u_0 a}{D_0}-a\del_x V(x_i)\,,\\
    \mathrm{ln}\left[\frac{k_r^{-}(x_i)}{k_r^{+}(x_i)}\right]&\approx& -\frac{u_0 a}{D_0}-a\del_x V(x_i)\,.
  \end{eqnarray}
\end{numparts}

Putting these together and replacing the sum with an integral, we obtain the entropy production rate from the active motion as
  \begin{eqnarray}
   \ac{\sigma} &=& \sum_{n^-,n^+}\int \mathrm{d}x\, p(x,n^-,n^+)\left[\frac{u_0^2 n^+}{D_0}-2n^-u_0\del_x V(x)+D_0 n^+ \del_x V(x)\right]\cr
    \cr
    &\phantom{}& +\sum_{n^-,n^+}\int \mathrm{d}x\, \frac{\del_x p(x,n^+,n^-)}{p(x,n^+,n^-)} \left[-u_0n^- +D_0 n^+ \del_x V(x) \right]\cr
    \cr
   &=&\left\langle\frac{u_0^2 n^+}{D_0}\right\rangle -2\left\langle n^-u_0\del_x V(x)\right\rangle\cr
   \cr
      &\phantom{}&+ \left\langle n^+D_0(\del_x V(x))^2\right\rangle - \left\langle n^+ D_0\del_x^2 V(x)\right\rangle\,,
      \label{eq:sigma_ac}
  \end{eqnarray}
     where the averages $\langle\dots\rangle$ are performed with regard to $p(x,n^-,n^+)$. 

Additionally, we allow for ordinary thermal translational diffusion with the current \cite{pietzonka_entropy_2017}
\begin{equation}
  j_{\mathrm{tr}} =-D_{\mathrm{tr}}p(x_i)\del_x\left[V(x_i)+\mathrm{ln}p(x_i)\right]\, ,
  \label{eq:j_tr}
\end{equation}
where $D_{\mathrm{tr}}$ is the thermal diffusion constant. For this active system, the entropy production stemming from thermal translation does not vanish and is given by 
  \begin{equation}
    \sigma_{\mathrm{tr}}= -\sum_{n^-,n^+}\int \mathrm{d}x \, j_{\mathrm{tr}}(x)\nabla V(x) = D_{\mathrm{tr}}\left\langle \del_x^2 V(x) - \left(\partial_x V(x)\right)^2\right\rangle\,,
    \label{eq:sigma_tr}
  \end{equation}
  where the second equality follows after a partial integration. 
\subsection{Activation statistics}
Since the activation and deactivation of reaction sites is a stochastic process, we now turn to the statistics of $n^-$ and $n^+$, which we have so far treated as fixed values. The probability of finding a single site either activated or deactivated is independent of the state of the other sites and identically distributed for all sites. Thus, the probability of finding $n^{l,r}$ active sites is given by a binomial distribution $B(n^{l,r}|N,p_A)$. For large $N$, this is well approximated by a Gaussian $\mathcal{N}(\mu,\sigma^2)$ with mean $\mu$ and variance $\sigma^2$, such that
  \begin{equation}
    p(n^{l,r}) \approx \mathcal{N}(Np_A,Np_A(1-p_A))\, ,
  \end{equation}
The probability distributions for $n^-$ and $n^+$, which are the difference and sum of $n^l$ and $n^r$, respectively, are then also Gaussian. For $n^-$ ($n^+$) the mean is subtracted (added) and the variance is the sum of the variance of $n^l$ and $n^r$, i.e.,

\begin{numparts} \label{eq:1d_prob_dist}
\begin{eqnarray}
  p(n^-) &\approx& \mathcal{N}(0,2Np_A(1-p_A))\,,\\ 
    p(n^+) &\approx& \mathcal{N}(2Np_A,2Np_A(1-p_A))\,.
\end{eqnarray}

\end{numparts}
In the limit $N\rightarrow\infty$, these approximations are exact. From (\ref{eq:1d_prob_dist}), we see that $n^+$ scales linearly with the total number of sites, whereas $n^-$ does not. However, in order to recover AOUP-statistics, we require that, for large $N$, the contribution to the dynamics from the terms proportional to $n^\pm$ in (\ref{eq:jac}) should be of the same order and independent of $N$. This implies that $Np_A$ does not depend on $N$ and is $\mathcal{O} (1)$, which we enforce by scaling $p_A\propto 1/N$. The constant $Np_A$ is the mean number of activated sites $\bar{n}_A$. Thus, we choose $p_A=\bar{n}_A/N$, where $\bar{n}_A$ will turn out to be a crucial parameter in the model. Physically, this rescaling of the activation probability ensures that the fluctuations around $0$ in $n^-$, which lead to the active propulsion, are not suppressed as the number of sites increases. At the same time, it ensures that the diffusive contribution to the active current stays finite. The rates $\omega_{\mathrm{on},\mathrm{off}}$ from (\ref{eq:def:omega}) can then also be expressed through $\bar{n}_A$ as 
\begin{equation}
  \omega_{\mathrm{on}} = \kappa \sqrt{\frac{\bar{n}_A/N}{1-\bar{n}_A/N}} \quad \quad\mathrm{and} \quad\quad \omega_{\mathrm{off}} = \kappa \sqrt{\frac{1-\bar{n}_A/N}{\bar{n}_A/N}}\,.
\end{equation}

Since the dynamics of $n^\pm$ is an equilibrium process, the distribution $p(n^\pm)$ can also be interpreted as a Boltzmann distribution $p(n^\pm)\sim \mathrm{exp}\left[-U_{n^\pm}(n^\pm)\right]$ with some effective potential $U_{n^\pm}$. We have already established that $p(n^\pm)$ are Gaussian distributions, which implies that the effective potentials for $n^-$ and $n^+$ read
\begin{equation}
  U_{n^-}(n^-)=(n^-)^2/(4 \bar{n}_A)\, \quad \mathrm{and}\quad U_{n^+}(n^+)=(n^+-2\bar{n}_A)^2/(4 \bar{n}_A)\,,
  \label{eq:effpot}
\end{equation}
respectively. The centers of these harmonic potentials correspond to the mean value of the distribution $p(n^\pm)$ and their widths correspond to the variance. Since the Boltzmann distribution is the steady-state solution of the Fokker-Planck equation for an overdamped particle in a harmonic potential, we get the condition on the current
\begin{equation}
  0 = -\partial_{n^\pm}\left\{\mu_{n^\pm}\left[-\del_{n^\pm}U_{n^\pm}(n^\pm)\right]p(n^\pm)\right\}+D_{n^\pm}\del^2_{n^\pm} p(n^\pm)\equiv -\del_{n^\pm}j_{n^\pm}\, .
  \label{eq:FPn-}
\end{equation}
Here, $\mu_{n^\pm}$ and $D_{n^\pm}$ are an effective mobility and an effective diffusion coefficient, respectively, for this internal degree of freedom. Since these are effective coefficients, they are not related by the Einstein relation and can be determined independently. Both are related to the timescale at which $n^\pm$ change and the appropriate choice of $\mu_{n^\pm}$ and $D_{n^\pm}$ lets us express the Langevin equation, which we will derive later, in such a way that we can identify $n^-$ as the colored noise source canonically used in describing AOUPs. The rate at which $n^{\pm}$ change is based on the rates $\omega_{\mathrm{on}/\mathrm{off}}$ from our microscopic model. The exit rate from a state with $2\bar{n}_A$ active sites is the sum of the rate at which an inactive site activates and the rate at which an active site deactivates. The timescale $\tau$, at which changes to $n^\pm$ take place is then the inverse of this exit rate, i.e.
\begin{eqnarray}
  \tau &=& (1/2)\left[Np_A\omega_{\mathrm{off}}+(N-\bar{n}_A)(1-p_A)\omega_{\mathrm{on}}\right]^{-1}\cr
  \cr
  &\approx& (1/2)\left(2 \bar{n}_A\right)^{-1}\left(\kappa \sqrt{\frac{1-\bar{n}_A/N}{\bar{n}_A/N}}\right)^{-1}\,,
\end{eqnarray}
where we have inserted (\ref{eq:def:omega}) and dropped terms of order $N^{-2}$. Choosing
\begin{equation}
  \mu_{n^\pm}  =  (2\bar{n}_A)/\tau \quad \mathrm{and}\quad D_{n^\pm} = 1/(2\tau^2)\,,
  \label{eq:def_mu_D}
\end{equation}
will yield the correct Langevin equations, as we will show later.
\subsection{Total entropy production}
Putting together (\ref{eq:FPn-}) and the currents (\ref{eq:1d_curr_tot}) and (\ref{eq:j_tr}) we obtain the stationary Fokker-Planck equation for the probability $p(x,n^-,n^+)$ as
\begin{eqnarray}
  0 &=& \del_t p(x,n^-,n^+) \cr
  \cr
  &=& -\partial_x\left[\ac{j}(x,n^-,n^+)+j_{\mathrm{tr}}(x)\right] -\partial_{n^-}j_{n^-}(n^-)-\partial_{n^+}j_{n^+}(n^+)\cr
  \cr
  &=& -\partial_x\left[u_0n^- -D_0n^+\partial_x V(x)\right]p(x,n^-,n^+)\cr
  \cr
  &\phantom{}& +D_{\mathrm{tr}}\partial_x\left\{\left[\partial_x V(x)\right]p(x,n^-,n^+)\right\}+ \left[(D_0 n^+ + D_{\mathrm{tr}})\partial_x^2\right] p(x,n^-,n^+)\cr
  \cr
  &\phantom{}& +\partial_{n^-}\left[\mu_{n^-}(2\bar{n}_A)^{-1} n^-p(x,n^-,n^+)\right]+ D_{n^-}\partial_{n^-}^2 p(x,n^-,n^+)\cr
  \cr 
  &\phantom{}& + \partial_{n^+}\left[\mu_{n^+}(2\bar{n}_A)^{-1}(n^+-2\bar{n}_A)p(x,n^-,n^+)\right]\cr
  \cr
  &\phantom{}& + D_{n^+}\partial_{n^+}^2 p(x,n^-,n^+)\,.
  \label{eq:FP_total}
\end{eqnarray}
Since the activation and deactivation of reaction sites is an equilibrium process, uncoupled from the dynamics in $x$, it does not produce entropy. The total entropy production is then the sum of (\ref{eq:sigma_ac}) and (\ref{eq:sigma_tr}), which reads
\begin{eqnarray}
  \sigma_{\mathrm{tot}} = \sigma_{\mathrm{ac}}+\sigma_{\mathrm{tr}}=&\phantom{}&\left\langle \frac{u_0^2 n ^+}{D_0}\right\rangle-2\langle u_0 n^- \del_x V(x)\rangle \cr
  \cr
  &\phantom{}&+D_0\langle n^+ \left[\del_x V(x)\right]^2   +n^+ \del_x^2 V(x)\rangle\cr
  \cr
  &\phantom{}&+ D_{\mathrm{tr}}\left\langle \del_x^2 V(x) - \left[\del_x V(x)\right]^2\right\rangle\,.
  \label{eq:sig_tot_1}
\end{eqnarray}
By multiplying (\ref{eq:FP_total}) with $V(x)$ and integration by parts, we obtain
\begin{eqnarray}
  D_{\mathrm{tr}}&\phantom{}&\left\langle\left[\partial_x V(x)\right]^2 - \partial_x^2 V(x) \right\rangle =\cr
  \cr
  &\phantom{}& \left\langle \partial_x V(x)\left[u_0n^--D_0n^+\partial_x V(x)\right]\right\rangle + D_0\left\langle n^+\partial_x^2 V(x)\right\rangle\, ,
\end{eqnarray}
which we plug into (\ref{eq:sig_tot_1}) to obtain 
\begin{equation}
  \sigma_{\mathrm{tot}} = \left\langle\frac{u_0^2 n^+ }{D_0}\right\rangle-u_0\left\langle n^-\partial_x V(x)\right\rangle\, .
  \label{eq:sigtot}
\end{equation}
This mean total entropy production rate in the steady state for this active particle constitutes our first main result.

\section{Comparing different measures for the departure from equilibrium} \label{sec:comparison}
To show that our model is equivalent to the AOUP model, we start by briefly introducing the canonical description of AOUPs \cite{caprini_entropy_2019,dabelow_irreversibility_2019,fodor_how_2016,fodor_irreversibility_2022,mandal_entropy_2017}, where we adopt the conventions from \cite{dabelow_irreversibility_2019,dabelow_irreversibility_2021} to our notation, which then reads
\begin{numparts}\label{eq:Langevin_old}
\begin{eqnarray}
  \dot{x}(t) &=& -D\del_x V(x(t),t) + \sqrt{2D_a}\eta(t) +\sqrt{2D}\xi (t)\,,\\
  \tau_a\dot{\eta}(t) &=& - \eta(t) + \zeta(t)\,,
\end{eqnarray}
\end{numparts}
where $\zeta(t)$ and $\xi(t)$ are independent unit white noises, by which we mean that they are correlated as $\left\langle\zeta(t)\zeta(t^\prime)\right\rangle = \left\langle\xi(t)\xi(t^\prime)\right\rangle = \delta(t-t^\prime)$. While early papers \cite{fodor_how_2016,mandal_entropy_2017} have not considered thermal noise $\xi(t)$, it has been included in more recent publications \cite{caprini_entropy_2019,dabelow_irreversibility_2019,fodor_irreversibility_2022}. The diffusion coefficient $D$ is equivalent to the thermal diffusion coefficient $D_{\mathrm{tr}}$ in our model. The active diffusion coefficient $D_a$ measures the amplitude of the Ornstein-Uhlenbeck process $\eta(t)$, which has correlations
\begin{equation}
  \langle\eta(t)\eta(t^\prime)\rangle = \left(1/2\tau_a\right)\mathrm{exp}\left(-|t-t^\prime|/\tau_a \right)\,,
\end{equation}
with a correlation time $\tau_a$. 

The Langevin equation corresponding to our model follows with (\ref{eq:def_mu_D}) from the Fokker-Planck equation (\ref{eq:FP_total}) as
\begin{numparts}    \label{eq:Langevin_all}
\begin{eqnarray}
  \dot{x}(t) = - D_{\mathrm{eff}}(t)\partial_x V(x(t)) + u_0n^-(t) + \sqrt{2D_{\mathrm{tr}}}\zeta_{\mathrm{tr}}(t) +\sqrt{2n^+ D_0}\zeta_{\mathrm{ac}}(t)\,,\\
  \cr
  \tau\dot{n}^-(t) = -n^-(t)+\zeta_{n^-}(t)\,,\\
  \cr
  \tau\dot{n}^+(t) = -\left(n^+(t)-2\bar{n}_A\right)+\zeta_{n^+}(t)\,,
\end{eqnarray}
\end{numparts}
with
\begin{equation}
  D_{\mathrm{eff}}(t)\equiv D_{\mathrm{tr}}+D_0 n^+(t)\, .
  \label{eq:def_Deff}
\end{equation} 
The stochastic processes $\zeta_{n^+}$, $\zeta_{n^-}$, $\zeta_{\mathrm{ac}}$ and $\zeta_{\mathrm{tr}}$ are all independent unit white noises. The variable $n^-$ follows an Ornstein-Uhlenbeck process and plays the role of the active noise source, while $u_0$ measures its amplitude and is equivalent to $\sqrt{2D_a}$. There is, however, a difference compared to the conventional description in (\ref{eq:Langevin_old}), which is the additional degree of freedom $n^+$. It enters the dynamics of $x(t)$ in two places. First, there is an additional noise source from the active process, which has an amplitude proportional to $n^+$. It arises because each active site contributes to diffusion and thus increases the active noise amplitude. This new noise source leads to an effective diffusion coefficient, which can be seen as active mobility. Both of these additional phenomena also appear in the microscopic model for ABPs \cite{pietzonka_entropy_2017}. Importantly, the fluctuations of $n^-$, which lead to the self-propulsion, and the amplitude of $n^+$ are of the same order, that is, as the self-propulsion speed increases, so does diffusion. 

With these Langevin equations, we can now compare our result for the total entropy production, (\ref{eq:sigtot}), to results already established in the literature, based on (\ref{eq:Langevin_old}). Calculating the entropy production of AOUPs on the level of Langevin dynamics relies on the time reversal of the trajectory, in which the reversal of the noise history can be chosen either even or odd. By choosing an odd time reversal, one implicitly also reverses the director of the particle, which is in conflict with the physical entropy production \cite{pietzonka_entropy_2017}. In our model, the path weight for a trajectory $x(t)$, given a trajectory of $n^-(t)$ and $n^+(t)$, reads
\begin{eqnarray}
    p\left[x(t) |n^-(t),n^+(t)\right]\sim  \mathrm{exp}\int_0^T\mathrm{d}t\,&\phantom{}& \left\{-\frac{1}{4 D_{\mathrm{eff}}}\left[\dot{x}-u_0n^-+D_{\mathrm{eff}}\del_x V(x)\right]^2\right.\cr
    \cr
    &\phantom{} &\left.+\frac{1}{2}D_{\mathrm{eff}}\del_x^2 V(x)\right\}\,,
\end{eqnarray}
where we omit the explicit time-dependence on the right-hand side for a more compact notation. The entropy production obtained from comparing the forward and backward probability under the time reversal $\tilde{x}(t)=x(T-t)$, $\tilde{n}^\pm(t)=n^\pm(T-t)$, denoted by $\mathcal{S}^+$ in \cite{fodor_irreversibility_2022}, reads 
\begin{eqnarray}
    \mathcal{S}^+ &=& T^{-1}\left\langle \mathrm{ln}\frac{p\left[x(t)|n^-(t),n^+(t)\right]}{p\left[\tilde{x}(t)|\tilde{n}^-(t),\tilde{n}^+(t)\right]} \right\rangle = \left\langle\frac{\dot{x}u_0n^-}{D_{\mathrm{eff}}}\right\rangle - \langle \dot{x}\del_x V (x)\rangle \cr
    \cr
    \cr
    &=& \left\langle \frac{u_0^2(n^-)^2}{D_{\mathrm{eff}}}\right\rangle - u_0 \langle n^-\del_x V(x) \rangle\,.
  \label{eq:S_plus}
\end{eqnarray}
For a comparison between $\sigma_{\mathrm{tot}}$ and $\mathcal{S}^+$, we use 
\begin{equation}
  \left\langle \frac{u_0^2(n^-)^2}{D_{\mathrm{eff}}}\right\rangle < \left\langle \frac{u_0^2(n^-)^2}{D_0n^+}\right\rangle < \left\langle \frac{u_0^2(n^+)^2}{D_0 n^+}\right\rangle = \left\langle \frac{u_0^2 n^+}{D_0}\right\rangle\,,
\end{equation}
due to the fact that $D_{\mathrm{tr}}>0$ from (\ref{eq:def_Deff}) and $n^{l,r}>0$. Hence, it follows that $\mathcal{S}^+<\sigma_{\mathrm{tot}}$. As discussed in \cite{pietzonka_entropy_2017} for ABPs, part of the underestimation comes from the fact that the Langevin description of the AOUP does not distinguish whether a jump occurs due to thermal or active noise, which both contribute differently to entropy production. In addition, the Langevin description does not distinguish whether the active noise causes a fuel-synthesizing or fuel-consuming reaction, which further increases the difference between these two measures of irreversibility.

In regimes where the energy exchanged with the active bath is much larger than the energy scale of the thermal bath, i.e., for $n^+ D_0\gg D_{\mathrm{tr}}$, setting $\zeta_{\mathrm{tr}}=0$ and discarding thermal noise is a tempting approximation. While this simplification is still thermodynamically consistent in our model, even in this case the entropy production $\mathcal{S}^+$ underestimates the total entropy production $\sigma_{\mathrm{tot}}$. In contrast, the active noise $\zeta_{\mathrm{ac}}$ is crucial for thermodynamic consistency in our model, as it ensures micro-reversibility. However, it has not been included in the canonical description (\ref{eq:Langevin_old}).

In our model, a putative time reversal that involves an odd reversal of the noise history would read $\tilde{n}^-(t)= -n^-(T-t)$, such that
\begin{eqnarray}
  \mathcal{S}^- \equiv T^{-1}\left\langle \mathrm{ln}\frac{p\left[x(t)|n^-(t),n^+(t)\right]}{p\left[\tilde{x}(t)|\tilde{n}^-(t),\tilde{n}^+(t)\right]} \right\rangle = \left\langle u_0 n^- \del_x V(x)\right\rangle\,.
\end{eqnarray}
The fact that this measure vanishes for $V=0$ highlights that this quantity cannot measure a physical entropy production \cite{pietzonka_entropy_2017}. Furthermore, it is clear that in the present model $n^\pm$ are necessarily even variables for a sensible time-reversal which can be interpreted as playing a recorded movie of the motion of the particle backwards. The interpretation given in \cite{dabelow_irreversibility_2019} that $\mathcal{S}^+$ measures the departure from equilibrium for a passive particle in an active environment and $\mathcal{S}^-$ for a self-propelled active particle can thus not be sustained in the present analysis.

\section{Generalization to \textit{d} dimensions}\label{sec:ddim}
So far, we have dealt with a particle moving in one dimension. However, our approach is generalizable to any number of dimensions, as we will show in this section. 
To this end, each reaction site $k$ is now associated with a director $\bn_k$, which points in the direction of forward propulsion if a reaction with rate $k^+$ takes place while activated. The quantity $n^-$ is then replaced by the quantity
\begin{equation}
  \bn^-\equiv\sum_{k\in\mathcal{A}} \bn_k\, ,
\end{equation}
which is the sum of all directors in the set of active sites, $\mathcal{A}$. This generalizes the definition from (\ref{eq:defnpm}), where the director for particles on the left is $n_k=+1$ and for particles on the right $n_k=-1$.
The diffusion amplitude $n^+$ is replaced by a tensor
\begin{equation}
  \bi{M}^+ \equiv \sum_{k\in\mathcal{A}} \bn_k \otimes \bn_k\, .
\end{equation}
The quantity $u_0 \bn^-$ still has the interpretation of a mean propulsion velocity for a given internal state. The internal state of the particle is now given by $\left\{\alpha_k\right\}$, where $\alpha_k$ is the state of a reaction site $k$, which is either active or inactive. The probability distributions $p(n^\pm)$ are then replaced by multivariate distributions $p(\bn^-)$ and $p(\bi{M}^+)$, which depend on the shape of the particle and the distribution of reaction sites, which are so far arbitrary. 

To obtain AOUP statistics in higher dimensions, we can consider a spherical particle with a dense, uniform distribution of $N$ total reaction sites in the continuum limit. In this case, $\bn^-$ is a multivariate Gaussian distribution with mean $\langle\bn^-\rangle = \bi{\mu}_{\bn^-}=0$ and covariance matrix
\begin{equation}
  C_{ij} = \delta_{ij} N\pi^{-1}p_A (1-p_A)\,,
\label{eq:covariance}
\end{equation}
where $i$ and $j$ are Cartesian coordinates. Since $\bn^-$ is the sum of identically distributed and independent random variables $\bn_k$, it follows from the central limit theorem that the probability distribution for $\bn^-$ is Gaussian, if we sum over sufficiently many active reaction sites. Spherical symmetry implies that its mean is $0$ and that there are no cross correlations. The same argument applies to $p(\bi{M}^+)$, which is now a matrix normal distribution with the same among-row and among-column covariance matrix as in (\ref{eq:covariance}). However, since the product $\bn_k\otimes\bn_k$ has strictly positive values, the mean of $p(\bi{M}^+)$ is not $0$, but rather $\pi^{-1}\bar{n}_A\mathds{1}$. 
All the relations we have derived for the one-dimensional case remain valid upon replacing $n^-$ with $\bn^-$ and $n^+$ with $\bi{M}^+$. The total entropy production, (\ref{eq:sigtot}), thus reads
\begin{equation}
  \sigma_{\mathrm{tot}} = \frac{u_0^2}{D_0}\left\langle\mathrm{Tr}\left(\bi{M}^+\right)\right\rangle - u_0 \left\langle\bn^-\nabla V(x) \right\rangle\,.
\end{equation}
In particular, the corresponding Langevin equations read
\begin{numparts}
\begin{eqnarray}
  \dot{\bi{x}}(t) &=&  -\bi{D}_{\mathrm{eff}}(t)\nabla V(\bi{x}(t)) + u_0 \bn^-(t) + \sqrt{2D_{\mathrm{tr}}}\boldsymbol{\zeta}_{\mathrm{tr}}(t) + \boldsymbol{\xi}_{\mathrm{ac}}(t)\,,\\
  \cr
  \tau\dot{\bn}^-(t) &=& -\bn^-(t)+\boldsymbol{\zeta}_{n^-}(t)\,,\\
  \cr
  \tau\dot{\bi{M}}^+(t) &=& -\left(\bi{M}^+(t)-2\bar{n}_A\mathds{1}\right)+\boldsymbol{\zeta}_{n^+}(t)\,,
\end{eqnarray}
\end{numparts}
with diffusion coefficient $\bi{D}_{\mathrm{eff}}\equiv D_{\mathrm{tr}}\mathds{1}+D_0\bi{M}^+(t)$ and active noise $\langle\boldsymbol{\xi}_{\mathrm{ac}}(t)\otimes\boldsymbol{\xi}_{\mathrm{ac}}(t^\prime)\rangle = 2D_0\bi{M}^+(t)\delta(t-t^\prime)$. The further noise correlations are given by $\langle\boldsymbol{\zeta}_{\mathrm{tr}}(t)\otimes\boldsymbol{\zeta}_{\mathrm{tr}}(t^\prime)\rangle =\langle\boldsymbol{\zeta}_{n^-}(t)\otimes\boldsymbol{\zeta}_{n^-}(t^\prime)\rangle = \mathds{1}\delta(t-t^\prime)$ and $\left\langle(\boldsymbol{\zeta}_{n^+})_{ij}(t)(\boldsymbol{\zeta}_{n^+})_{kl}(t^\prime)\right\rangle =\delta_{ik}\delta_{jl}\delta(t-t^\prime)$.

\section{Conclusion} \label{sec:conclusion}
We have presented a microscopic Markovian model for a particle which leads to AOUP statistics in the continuum limit. For our model, the well-established tools of stochastic thermodynamics can easily be applied, which let us calculate the mean total entropy production in the steady state. From a Fokker-Planck equation, we have then constructed the corresponding Langevin equation, which additionally features a fluctuating diffusivity. We have then applied concepts developed for measuring the departure from equilibrium for AOUP dynamics and found that the physical entropy production in our microscopic model is always larger than pseudo-entropy productions inferred from Langevin trajectories alone. In addition, we do not have to deal with the intricacies of an ambiguous time-reversal. Even though we have treated only one active particle, it is straightforward to extend our approach to many AOUPs and to AOUPs interacting with passive particles following the lines of \cite{pietzonka_entropy_2017}, where this has been done for ABPs. The resulting thermodynamically consistent model for an active bath of AOUPs can then be used to study efficiency and alike in models of such generalized active heat engines. Furthermore, our analysis has revealed one possible mechanism leading to AOUP-like motion, which is a uniform distribution of many propulsion sites and a low activation probability of each site. The construction of thermodynamically consistent models underlying the effective dynamics of active particles can thus yield insight and serve as a benchmark for effective descriptions. These models bridge the gap between understanding the underlying, thermodynamically relevant, but inaccessible processes on the one hand and the observable degrees of freedom and effective measures for the departure from equilibrium on the other.

\section*{Acknowledgements}
We thank Julius Degünther and Jann van der Meer for insightful discussions.
\section*{References}
\bibliography{AOUP_final.bib}

\end{document}